# Debye relaxation in high magnetic fields

J.S. Brooks[1*], R. Vasic[1+], A. Kismarahardja[1],

E. Steven[1], T. Tokumoto[1], P. Schlottmann[1] and S. Kelly[2]

[1]Physics Department and National High Magnetic Field Laboratory, Florida State University, Tallahassee FL 32310 USA.

[2]Department of Chemistry, University of Alabama at Huntsville, Huntsville, AL, 35899 USA

Abstract

Dielectric relaxation is universal in characterizing polar liquids and solids, insulators, and semiconductors, and the theoretical models are well developed. However, in high magnetic fields, previously unknown aspects of dielectric relaxation can be revealed and exploited. Here, we report low temperature dielectric relaxation measurements in lightly doped silicon in high dc magnetic fields B both parallel and perpendicular to the applied ac electric field E. For B//E, we observe a temperature and magnetic field dependent dielectric dispersion $\varepsilon(\omega)$ characteristic of conventional Debye relaxation where the free carrier concentration is dependent on thermal dopant ionization, magnetic freeze-out, and/or magnetic localization effects. However, for B$\perp$E, anomalous dispersion emerges in $\varepsilon(\omega)$ with increasing magnetic field. It is shown that the Debye formalism can be simply extended by adding the Lorentz force to describe the general response of a dielectric in crossed magnetic and electric fields. Moreover, we predict and observe a new transverse dielectric response $E_H \perp B \perp E$ not previously described in magneto-dielectric measurements. The new formalism allows the determination of the mobility and the ability to discriminate between magnetic localization/freeze out and Lorentz force effects in the magneto-dielectric response.

*Corresponding Author:

J.S. Brooks, brooks@magnet.fsu.edu; Ph. 1-850-644-2836 (Fax: -5038)

+Present address: North Carolina State University, Department of Physics, Box 7518, 851 Main Campus Dr., Raleigh, NC 27695

## I. INTRODUCTION

Silicon is one of the most intensely studied materials due to its profound importance in electronic devices. Even with the promise of direct gap, high mobility materials such as gallium arsenide, or "flexible" organic-based transistors, silicon still holds the technological advantage industry wide. For these purposes, knowledge of the properties of silicon under ambient conditions of temperature, magnetic field, etc. are sufficient. However, at low temperatures and in high magnetic fields, technologically important materials and devices can yield surprises, such as the discovery of the quantum Hall effect in Si MOSFETs [1].

In 1929, Debye described a model[2] for the response of electric dipoles to an alternating electric field where the essential dynamics were described by the exponential relaxation of dipole polarization with time. This model lead to a very simple description of the complex dielectric constant, namely

$$\varepsilon(\omega) = \varepsilon_\infty + (\varepsilon_s - \varepsilon_\infty)/(1 + i\omega\tau). \qquad (1)$$

Here $\varepsilon(\omega)$ depends only on the zero and high frequency limits ($\varepsilon_s$ and $\varepsilon_\infty$) and the product of the frequency and relaxation time ($\omega\tau$). Refinements of the Debye model allow a distribution of relaxation times introduced through an empirical exponent $\gamma$, such as the Cole-Cole formula[3]:

$$\varepsilon(\omega) = \varepsilon_\infty + (\varepsilon_s - \varepsilon_\infty)/(1 + (i\omega\tau)^\gamma). \qquad (2)$$

This distribution leads to memory or retardation effects in the scattering process, e.g. an interference between scattering centers such that the processes are no longer independent. The Debye and Cole-Cole related models are applicable to the description of a variety of liquids[4], semiconductors[5], magnetic systems [6], and other dielectric materials including soils[7]. In practice, the frequency ω of the electric field and/or experimental parameters that affect the relaxation rate τ such as temperature or magnetic field can be used to characterize ε(ω) over a wide range of ωτ above and below the "resonant" condition ωτ = 1. In this report we describe dramatic changes in the form of equation (1) that appear in high magnetic fields in lightly doped silicon (N ~ $10^{14}$ cm$^{-3}$), as shown in both experiments and a complementary model. To be clear, we emphasize that the samples are insulating, very far away from the metal insulator transition, and the relevant frequency is nearly dc when compared with plasma frequency and/or infrared energy scales. Hence the only contribution of the electronic structure is through the thermal activation of carriers, and interband transitions, universal scaling, etc., are not relevant to the present work.

## II. EXPERIMENT

The dc dielectric constant $ε_s$ of pure crystalline silicon[8] is about 11.7 $ε_0$ where $ε_0$ is the permittivity of free space. The lead configuration used in this work is shown in Fig. 1a. Here commercial silicon wafers of thickness 0.5 mm were cleaved in to 5 x 5 mm squares and capacitive (and Hall) electrodes were made directly to the silicon surface (and edges) with silver or carbon paste. The ac electric field (of order 0.1 V/cm) was therefore always parallel to the (100) direction perpendicular to the plane of the wafer. Independent Hall measurements in the van der Pauw configuration yielded a carrier concentration of N ~ $10^{14}$cm$^3$ (i.e. well below the critical concentration of order $10^{18}$/cm$^3$ for the metal insulator transition[9]). The samples were mounted on a rotation probe in a helium cryostat in high magnetic fields. A standard capacitance bridge and lock-in amplifiers were used to determine the real and imaginary parts of the dielectric response. In all cases, the bridge frequency, magnetic field, and sample orientation (i.e. angle between the electric and magnetic fields) were fixed, and the real and imaginary bridge signals (in the linear response region) were recorded vs. temperature. To insure that the results reported herein were due to the bulk silicon material, and not from the silver paint electrodes, control experiments with a contact-less electrode configurations were employed to eliminate the possibility that Shottkey effects influenced the data. Likewise, several different samples in both B//E and B⊥E configurations were used in the experiments reported herein to verify reproducibility.

## III. RESULTS

A representative measurement relevant to the work described below is shown in Fig. 1b for a lightly doped silicon sample. Here the dielectric response is measured vs. temperature at constant frequency (without magnetic field) where the real and imaginary parts of $\varepsilon(\omega)$ (= $\varepsilon' + i\varepsilon''$) are plotted vs. temperature, and the Cole-Cole plot[3] for $\varepsilon'$ vs. $\varepsilon''$ is shown in the inset. As discussed below, it is the exponential dependence of $\tau$ on temperature through the carrier concentration n that allows $\omega\tau$ to range from $\omega\tau \ll 1$ to $\omega\tau \gg 1$, revealing the full range of dielectric behavior. The results indicate a typical Debye-like dielectric response with a finite distribution of relaxation rates described by the exponent $\gamma = 0.87$, where $\gamma < 1$ indicates that long time effects (small frequencies) play a role as is expected for glassy behavior. (We will return to the significance of $\gamma$ in Section IV below.)

In magnetic fields, there are dramatic changes in the dielectric relaxation behavior of lightly doped silicon, especially for the condition B⊥E, as shown in Fig. 2 where the Debye-like behavior (for B = 0 T) becomes more complex. With increasing field, the resonant condition (determined herein by the parameters associated with the peak in ε") moves to higher temperature, and anomalous dispersion appears, manifested by peaks and minima in ε', and the narrowing and increase in amplitude of the resonant peak structure in ε". Central to the main result of this work is that anomalous dispersion appears only when there is a finite component of the magnetic field perpendicular to the electric field. This is shown explicitly for ε'' in Fig. 3 where the sample is rotated in field from B//E to B⊥E.

Anomalous dispersion is a natural result of the driven, damped harmonic oscillator associated with a hydrogen-like bound carrier state[10]:

$$\varepsilon = \varepsilon_s + \varepsilon_s \omega_p^2 / (\omega_0^2 - \omega^2 + i\omega\alpha). \qquad (3)$$

Here $\omega_0$ is the oscillator resonant frequency, $\alpha$ is the damping factor proportional to the displacement velocity, and $\omega_p^2$ is the plasma frequency $Ne^2/\varepsilon_s m$. However, no set of parameters associated with equation (3) was capable of modeling the behavior associated with our experiments: the frequency scale for a harmonic oscillator associated with a hydrogenic donor state is of order 8 THz, whereas the relaxation rate in our experiments is in the range $10^4$ Hz or less (i.e. $\omega_0 \gg 1/\tau$). Moreover, equation (3) does not explicitly include the effects of magnetic field.

## IV. MODEL

To describe the anomalous dispersion in our experiments for B⊥E, the Debye relaxation problem must be augmented to properly include the magnetic field beyond simple magnetic freeze out and localization effects. We treat lightly doped silicon as a medium of ionized impurities where the resulting carrier density is thermally activated. In the present case, the electron ionization energy is of order 45 meV based on Arrhenius analysis of the condition ωτ =1 vs. T (as in Fig. 2)[11], corresponding very closely to the value for Si:B -acceptor or Si:P-donor states.[8] In an externally applied electric field **E**, there will be a relative displacement of positive and negative charge, and this will result in a polarization field which is proportional to the displacement **x**, as shown in Fig. 4. The important differences between equation (3) and our model are: 1) the restoring force is due to the polarization field, not to the hydrogen bound state; 2) the number of carriers that can be polarized depends on thermal activation; 3) the magnetic field is introduced explicitly through the Lorentz force; and 4) there is no acceleration term in the equations of motion, as we now describe.

The response of a test charge e to a Lorentz force is assumed to be in the diffusive regime, and hence no acceleration (mass dependent) term is present[6]. We also consider a simple Markovian process, and therefore do not include the slight deviation of γ from unity as indicated in the Cole-Cole analysis in Fig. 1b. (A treatment involving a fractal environment[12] could in principle be used to refine the model, but as shown below, the present model is adequate to describe the essential features of the experiment.) Assuming each test charge reacts to the mean field of the bulk charge displacement $E_p$ and to the externally applied Lorentz field, we may write a general expression for the motion of the charge associated with crossed electric and magnetic fields ($\mathbf{E} = (E,0,0)$ and $\mathbf{B} = (0,0,B)$) as:

$$e(dE_p/dx)x + e\alpha x' = eE + eBy' \quad \text{and} \quad e(dE_p/dy)y + e\alpha y' = -eBx' \qquad (4)$$

where the primes refer to the time derivatives ( the time dependence $e^{-i\omega t}$ is explicit in the x, y, and E terms) , the linear terms in displacement represent the restoring force associated with the polarization field, and $e\alpha$ is the damping factor associated with the velocity term. For simplicity, we divide equation (4) by e and let both $dE_p/dx$ and $dE_p/dy$ = $k = n/\varepsilon_0$ (from Fig. 4 above). Hence equation (4) becomes

$$kx + \alpha x' = E + By' \quad \text{and} \quad ky + \alpha y' = -Bx' . \qquad (5)$$

For E = 0 and B = 0, the equations reduce to the solution $x = x_0 e^{-t/\tau}$ (where $\tau = \alpha/k$), namely that an initial polarization $nx_0$ will relax exponentially to zero, which is the condition for Debye relaxation (similarly for y).[13] Note that $1/\alpha$ is the carrier mobility $\mu$, and since $\tau = \alpha/k = \varepsilon_s/n\mu = \varepsilon_s/\sigma = \varepsilon_s\rho$, $\tau$ is analogous to the relaxation time of a "RC circuit" (where $\sigma$ and $\rho$ are the conductivity and resistivity respectively). In the presence of an oscillating field E (where B = 0), the solution of equation (5) is $x = (E/k)*(1 – i\omega\tau)^{-1}$ which is identical to the frequency dependent relaxation associated with the Debye model. When both E and B are non-zero, the solutions of equation (5) for x and y (*where x and y now represent the time independent, phase information*) take the form:

$$x = (E/k)[(1-i\omega\tau) - (\omega B/k)^2/(1-i\omega\tau)]^{-1}$$

$$\text{and } y = i(E/k)(\omega B/k)[(1-i\omega\tau)^2 - (\omega B/k)^2]^{-1} \quad (6)$$

From equation (6) and the relationship $\varepsilon_x(\omega) = \varepsilon_s + Nex/E$ we may obtain the real and imaginary components of the dielectric response in the x (and E) direction in the form

$$\varepsilon_x' = \varepsilon_s + (Ne/k)C_1/(C_1^2 + (\omega\tau C_2)^2) \text{ and } \varepsilon_x'' = (Ne/k)\omega\tau C_2/(C_1^2 + (\omega\tau C_2)^2)$$

where $C_1 = [1 - (\omega B/k)^2/(1+(\omega\tau)^2)]$ and $C_2 = [1 + (\omega B/k)^2/(1+(\omega\tau)^2)]$. $\quad (7)$

In the y direction only the y-polarization is present, hence $\varepsilon_y(\omega) = Ney/E$ and

$$\varepsilon_y' = -2(Ne/k)\omega\tau(\omega B/k)/(R^2 + 4(\omega\tau)^2) \text{ and } \varepsilon_y'' = (Ne/k)(\omega B/k)R/(R^2 + 4(\omega\tau)^2)$$

where $R = (1-(\omega\tau)^2-(\omega B/k)^2)$. $\quad (8)$

Comparison of equations (7) and (8) with our experimental results can be best demonstrated in the temperature domain at fixed frequency. In the model, the temperature dependence is contained in the parameters $\alpha$ (= $1/\mu$) and k. Typically, for doped silicon, the mobility $\mu$ has a power law dependence[8] on T, and usually saturates or decreases at low T. However, since the charge density $n = n_0\exp(-E_a/kT)$, $k = (n_0/\varepsilon_s)\exp(-E_a/kT)$. This means the temperature dependence is primarily in the polarization field which depends on n through k. Using the approximate experimental parameters $n_0/\varepsilon_s =$ 1.3 10$^{14}$ V/m$^2$ (corresponding to N =7.3x10$^{15}$ carriers/cm$^3$ ), $E_a \sim$ 450 K and f = 30 KHz, (where we compute $\alpha$ from k/$\omega$ under the condition that at 20 K, $\omega\tau = 1$, and take the mobility as temperature independent) we may plot equation (7) vs. T in Fig. 5a and 5b. As in the experiments, with increasing magnetic field anomalous dispersion appears in $\varepsilon_x'/\varepsilon_s$, and the peak for $\varepsilon_x''/\varepsilon_s$ narrows, increases in amplitude, and moves to higher temperatures. It is clear that the model captures the essential features of the data for $\varepsilon_x(\omega)$ when compared with for instance the behavior in Fig. 2.

Of equal importance is that our model predicts a transverse Hall response (for E//x, and B//z, there will be a voltage in the y-direction), as given in equation (8), and this is shown in Fig. 5c and 5d. To test the prediction of the model for $\varepsilon_y(\omega)$, we performed a set of measurements where the voltage due to the transverse response was measured simultaneously with the longitudinal response, and data were taken for forward (B+) and reverse (B-) magnetic fields. The transverse components of the signals were then obtained by subtracting the B+ and B- traces. The results are presented in Fig. 6. It is evident that, as in normal Hall effect measurements, the longitudinal and transverse components are mixed due to lead misalignment. However, there is no question that the derived transverse signal is exactly as predicted for the $\varepsilon_y(\omega)$ response in equation (8). To our knowledge, no theoretical and experimental treatment of the transverse dielectric response for B$\perp$E has previously been reported.

## V. DISCUSSION

The model described above for the case B$\perp$E gives access to important physical parameters such as the mobility. At zero magnetic field where the dielectric response depends solely on $\omega\tau$, the temperature and frequency dependence of the peak (where $\omega\tau = 1$) in $\varepsilon_x''$ yields $E_a$ in the range where the mobility does not vary significantly over the experimental range of temperature. Since the carrier concentration depends exponentially on T, whereas the mobility has a power law T-dependence, this is easily satisfied, particularly at lower temperatures. In finite magnetic field for B⊥E, the peak in $\varepsilon_x''$ does not correspond to $\omega\tau = 1$, but now depends on $\omega B/k$. This is clearly demonstrated in Fig. 3 where the anomalous dispersion peak shifts to higher temperature for B⊥E. The relationship between $\omega\tau$ and $\omega B/k$ for the peak condition $d\varepsilon_x''/dT = 0$ is most easily obtained from the corresponding condition $d\varepsilon_y'/d(1/k^2) = 0$ (where we note that N ~ k), which yields the resonant condition for each peak (vs. frequency, field, and temperature):

$$(\omega\tau)^2 + (\omega B/k)^2 = 1 \quad \text{(at resonance)}. \qquad (9)$$

Equation (9) is very useful, since over a narrow range of temperature, the longitudinal mobility $\mu_x$ may be obtained from the relationship

$$(\omega B/k)/(\omega\tau)/B = 1/\alpha = \mu_x. \qquad (10)$$

From equations. (9) and (10), $\mu_x$ can be determined by a simple rotation experiment as described in Fig. 3. This is accomplished by comparing the data at 8 T for B//E which is described by equation (1) (where $\omega\tau$ is field dependent, but unity at the peak in $\varepsilon_x''$), with the data for B⊥E as described by equation (7), where the resonant peak now occurs for the conditions in equation (9). Since $\omega\tau$ does not depend on the field orientation, we can use equation (1) to compute the value of $\omega\tau$ at the position of the shifted B⊥E peak, using the B//E data. We can then obtain $\omega B/k$ from equation (9), and $\mu$ from equation (10), which in this case yields $\mu_x \sim 8540$ cm$^2$/Vs.

The transverse mobility $\mu_y$ may also be obtained from $\varepsilon_y'$ (equation (8)) at resonance, since it reduces to $\varepsilon_y'(R=0) = -(Ne/k)B\mu/2$, or since $Ne/k = \varepsilon_s - \varepsilon_\infty$ ,

$$\mu_y = -2\,\varepsilon_y'(R=0)/(\,(\varepsilon_s-\varepsilon_\infty)\,B). \qquad (11)$$

We note here that some care is necessary since the transverse signal is derived for the electric field, and experimentally a voltage is measured. Hence the sample and contact geometry must be carefully taken into account.

The magnetoresistance due to magnetic localization and/or magnetic freeze-out may also be obtained from the condition B//E using conventional methods [14]. However, for B⊥E, the magnetoresistance analysis is more complicated due to the additional contribution of the Lorentz force, as noted in the discussion above for $\mu_x$ based on the analysis of Fig. 3.

## VI. SUMMARY

In summary, in lightly doped silicon, due to the activated dependence of the ionized carrier concentration on temperature, the Debye "resonant" signature in the dielectric response is accessible over a broad range of temperature and frequency. (1) For zero magnetic field, the activation energy of the carrier density is obtained from the frequency and temperature dependence through the condition $\omega\tau = 1$. (2) For B//E, the magnetoresistance may be found at constant temperature from the frequency and field dependence, where these effects arise from magnetic localization at lower fields and magnetic freeze out at higher fields[14]. In the Debye model, all of these effects enter through the temperature and field dependence of the carrier concentration, and to a lesser degree through the carrier mobility, both of which influence the relaxation time $\tau$. An advantage of high magnetic fields is the ability to resolve different carrier sites, due to different resonant conditions. (3) For $B \perp E$, there is the additional effect due to the parameter $\omega B/k$, and we here provide a complete theoretical description of the dielectric response under this condition for both longitudinal and transverse directions. It is from the $B \perp E$ field dependence that the longitudinal mobility $\mu_x$ can be obtained from equation (10), and the transverse (Hall) mobility $\mu_y$ from equation (11).

## ACKNOWLEDGEMENTS

Work supported in part by NSF DMR 0602859, and performed at the NHMFL which is supported by NSF Cooperative Agreement No. DMR-0084173, the State of Florida, and the DOE. PS is supported by the DOE under grant Nr. DE-FG02-98ER45707.

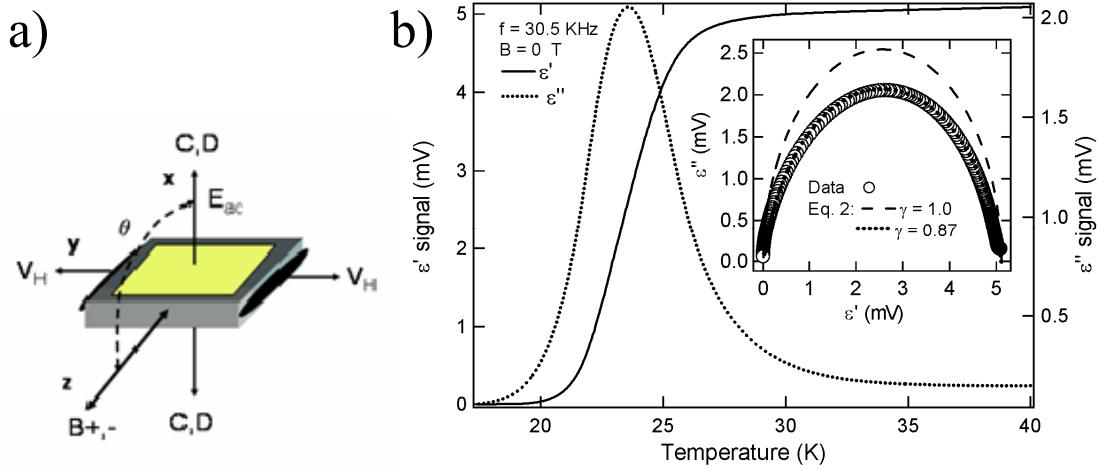

Figure 1: (Color online) Sample configuration and temperature dependent dielectric response. (a) Lead configurations for the sample used in this work. C,D are the capacitive leads in the direction of the ac electric field $E_{ac}$, and the $V_H$ leads are the transverse contacts. The reversible dc magnetic field is along z for B⊥E (shown in figure for θ = 90 degrees), and along x for B//E (θ = 0 degrees). (b) Dielectric response of silicon sample #1 vs. temperature for $\omega/2\pi$ = 30.5 KHz at zero magnetic field. Near T = 24 K, $\omega\tau \sim 1$ ($\omega\tau$ decreases for increasing T). Inset: Cole-Cole plot for ε" vs. ε'. The best fit of equation (2) to the data is for γ = 0.87.

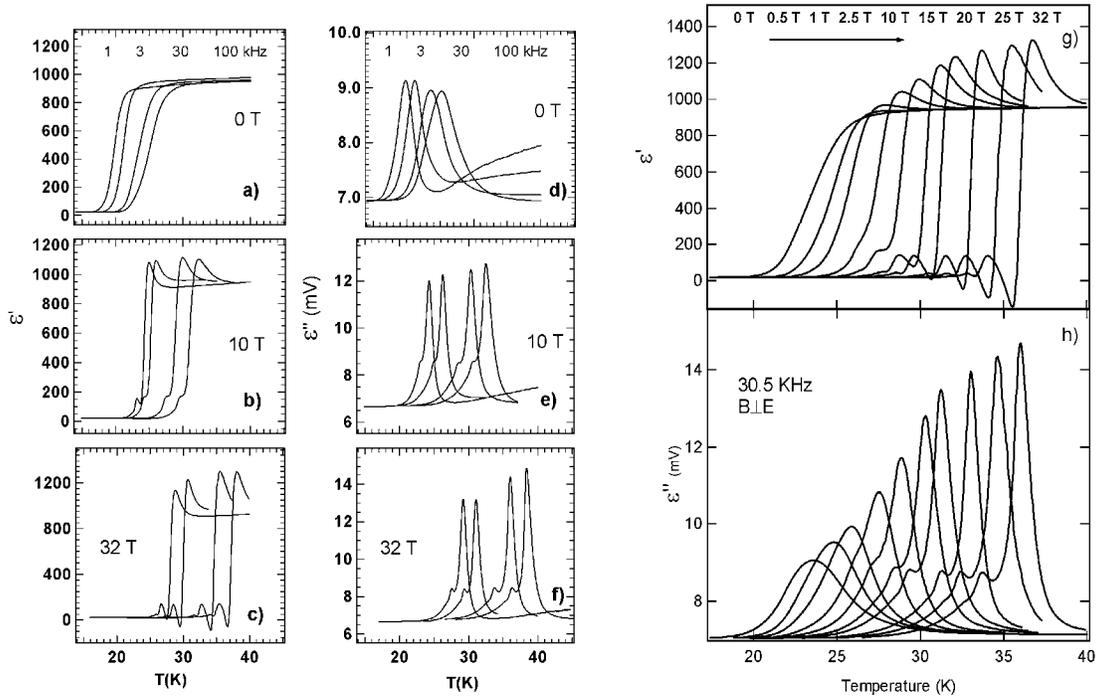

Figure 2. Temperature dependence of the real ($\varepsilon'$) and imaginary ($\varepsilon''$) dielectric signals for silicon sample #1 for B$\perp$E. (a-f) Frequencies of 1, 3, 30, and 100 kHz at constant magnetic fields of 0, 10, and 32 T. Deviations in the low frequency response, e.g. Fig. 2d, arise from glassy behavior[15]. (g-h) Magnetic field dependence of $\varepsilon'$ and $\varepsilon''$ at constant frequency from 0 to 32 T. At high fields additional small resonances from minority dipolar states can be resolved.

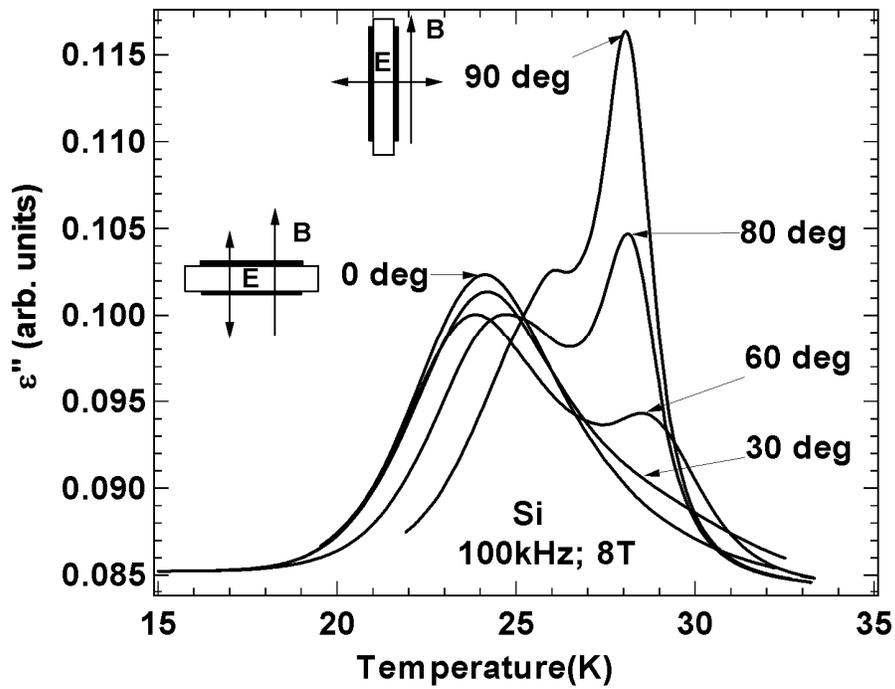

Figure 3. Angular dependence of ε" for silicon sample #1 at 100 kHz and 8 T. The anomalous dispersion appears when there is a finite component of B perpendicular to E. (0 degrees: B//E; 90 degrees: B⊥E).

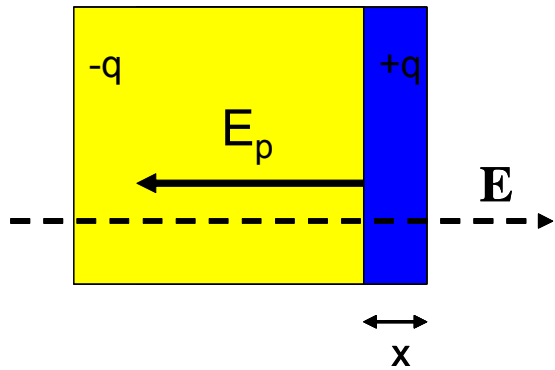

Figure 4. (Color online) Model for the dielectric response. A displacement of charge ± q due to an external field E in a dielectric produces a polarization field $E_p$ which increases linearly with the displacement x. For a charge density n (= Ne), the polarization charge per unit area is nx, and the polarization field is therefore $E_p = nx/\varepsilon_s$. As discussed in the text, in the absence of a driving field, an initial displacement will relax exponentially to zero with a time constant $\tau$.

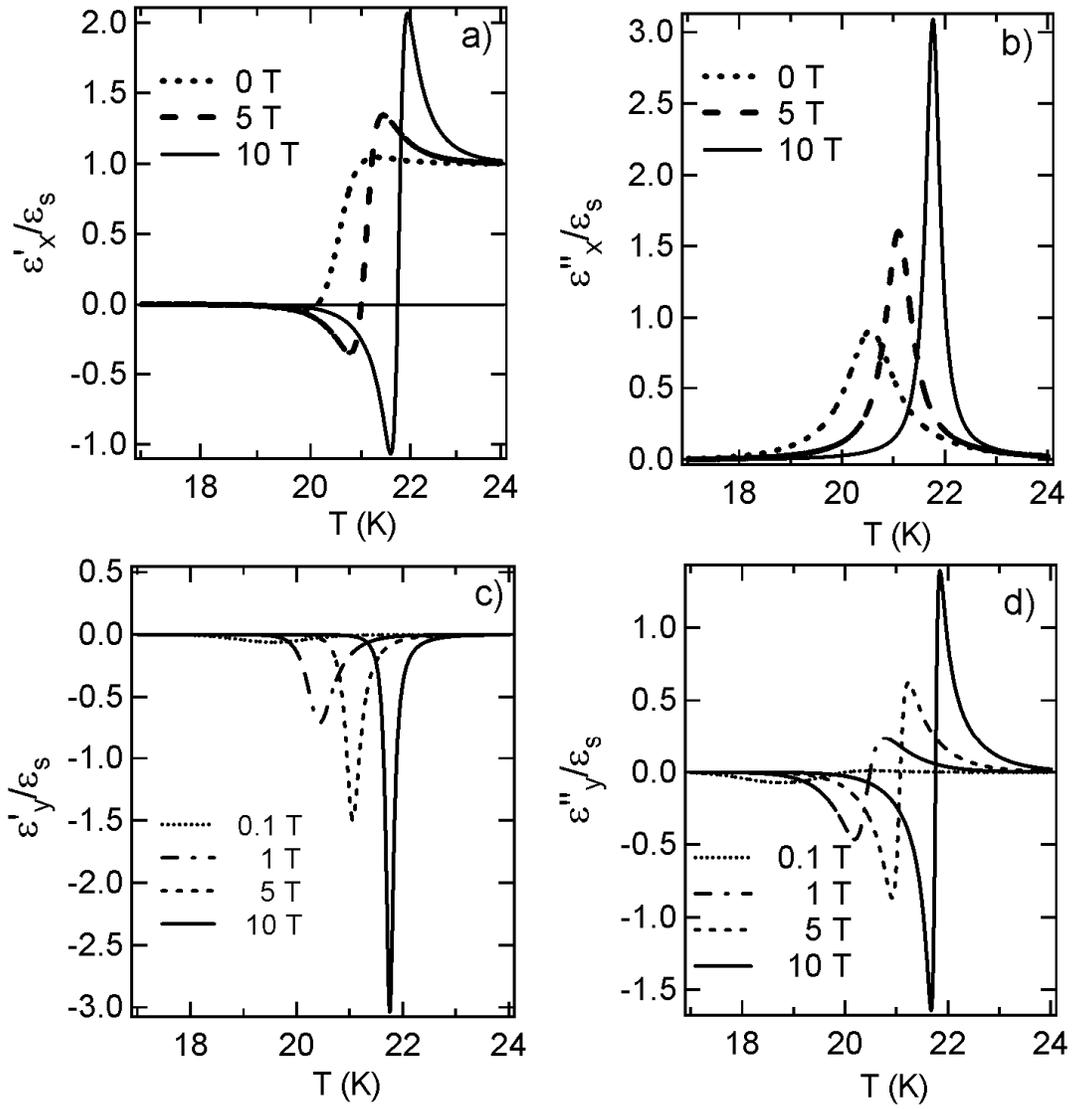

Figure 5. Predictions of the model. (a) $\varepsilon_x'/\varepsilon_s$ and (b) $\varepsilon_x''/\varepsilon_s$ vs. temperature from equation (7) at different magnetic fields for B⊥E. (c) $\varepsilon_y'/\varepsilon_s$ and (d) $\varepsilon_y''/\varepsilon_s$ vs. temperature at different magnetic fields for B⊥E from equation (8). Transverse signals only appear for B ≠ 0

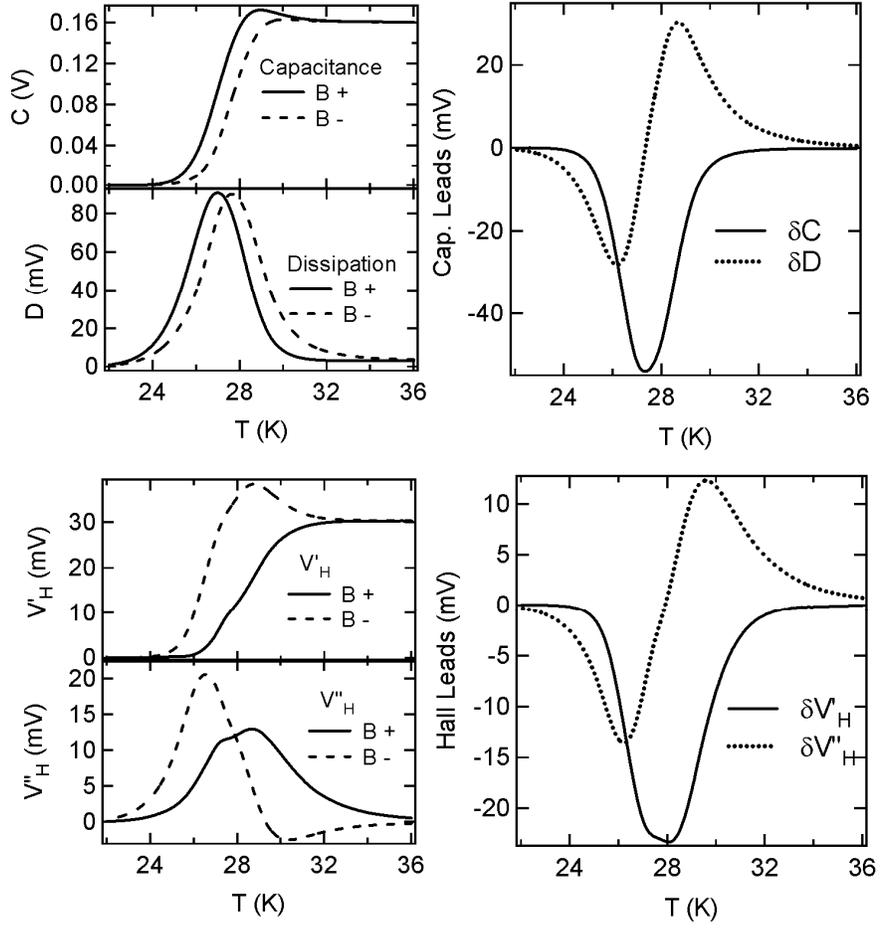

Figure 6. Longitudinal (C, D) and transverse ($V_H'$, $V_H''$) dielectric response for $B \perp E$ at 10 KHz and 1.23 T for sample # 2. Left panels: capacitance and Hall signals for forward and reverse field. Right panels: Difference signals showing transverse components (compare with equation (8) and Fig. 5). The signs for curves $\delta C$ and $\delta D$ have been reversed to account for the 180° ambiguity in the phase of the lead configuration.